\documentclass{article}

\PassOptionsToPackage{numbers}{natbib}

\usepackage{bbding}
\usepackage{enumitem}

    \usepackage[final]{neurips_data_2023}

\usepackage[utf8]{inputenc} 
\usepackage[T1]{fontenc}    
\usepackage{hyperref}       
\usepackage{url}            
\usepackage{booktabs}       
\usepackage{amsfonts}       
\usepackage{nicefrac}       
\usepackage{microtype}      
\usepackage[table,xcdraw]{xcolor}
\usepackage{graphicx}
\usepackage{caption}

\title{Berkeley Open Extended Reality Recordings 2023 (BOXRR-23): 4.7 Million Motion Capture Recordings from 105,852 Extended Reality Device Users}

\author{%
  Vivek Nair \\
  UC Berkeley\\
  Berkeley, CA 94720 \\
  \texttt{vcn@berkeley.edu} \\
  \And
  Wenbo Guo \\
  UC Berkeley\\
  Berkeley, CA 94720 \\
  \texttt{henrygwb@berkeley.edu} \\
  \And
  Rui Wang \\
  UC Berkeley\\
  Berkeley, CA 94720 \\
  \texttt{ruiwang813@berkeley.edu} \\
  \AND
  James F. O'Brien \\
  UC Berkeley\\
  Berkeley, CA 94720 \\
  \texttt{job@berkeley.edu} \\
  \And
  Louis Rosenberg \\
  Unanimous AI\\
  Pismo Beach, CA 93448 \\
  \texttt{louis@unanimous.ai} \\
  \And
  Dawn Song \\
  UC Berkeley\\
  Berkeley, CA 94720 \\
  \texttt{dawnsong@berkeley.edu} \\
}

\begin{document}

\maketitle

\begin{abstract}
Extended reality (XR) devices such as the Meta Quest and Apple Vision Pro have seen a recent surge in attention, with motion tracking ``telemetry'' data lying at the core of nearly all XR and metaverse experiences.
Researchers are just beginning to understand the implications of this data for security, privacy, usability, and more, but currently lack large-scale human motion datasets to study.
The BOXRR-23 dataset contains 4,717,215 motion capture recordings, voluntarily submitted by 105,852 XR device users from over 50 countries.
BOXRR-23 is over 200 times larger than the largest existing motion capture research dataset
and uses a new, highly-efficient and purpose-built XR Open Recording (XROR) file format.
\end{abstract}

\section{Introduction}
\label{intro}

For decades, human motion capture (MoCap) recordings have been an important resource in a variety of fields, ranging from animation and computer-generated imagery (CGI) to authentication and human-computer interaction (HCI). Recently, the proliferation of extended reality (XR) devices has created a prominent new application for this data, with motion data being central to almost all XR and ``metaverse'' experiences. Since 2002, at least 25 motion capture datasets have been created based on laboratory studies of up to a few hundred users to facilitate research in this important domain.

An emerging area of interest for security and privacy researchers is the passive identification and authentication of XR users based on their movement patterns. Until recently, XR identification and authentication studies have been limited to a few hundred users due to the lack of large-scale human motion datasets. By contrast, studies involving traditional biometrics, such as fingerprints or facial recognition, typically use datasets involving 100,000 or more subjects \cite{nist}.

In this paper, we introduce the BOXRR-23 dataset, which contains 4,717,215 motion capture recordings uploaded by 105,852 XR device users from over 50 countries. Our data is derived from two popular VR games, ``Beat Saber'' and ``Tilt Brush.'' In addition to being more diverse and ecologically valid than laboratory studies, BOXRR-23 is over 200 times larger than the largest known public motion capture dataset. We recently used this dataset, for the first time, to demonstrate that XR motion data provides a biometric signal on par with fingerprints \cite{nair2023unique}. The identification result, published in \textit{USENIX Security '23}, was made possible by this novel dataset. Moreover, we envision the potential uses of this data may go far beyond security and privacy to include areas such as motion synthesis, human-computer interaction, and machine learning research.

In addition to assembling this dataset from three public sources and enriching it with additional metadata, we developed a new ``Extended Reality Open Recording'' (XROR) file format due to the lack of an existing standard format suitable for this use case. The XROR format is about 30\% more space efficient than the original file formats, without loss of precision.

To help interested researchers evaluate this dataset, we provide documentation pursuant to a number of open standards, including Datasheets for Datasets \cite{gebruDatasheetsDatasets2020} and Dataset Nutrition Labels \cite{holland2018dataset}.
Furthermore, we conducted a large-scale survey ($N=1,006$) of the users contained in this dataset to better understand their demographics, the results of which are summarized herein.

\section{Background}
\label{background}

Since the 1990s, computerized motion tracking systems have been used for animation and CGI in a large number of popular movies, television series, and video games. A typical commercial motion capture solution uses optical tracking or inertial measurement units (IMUs) to measure the location of various body parts, with prices ranging from \$10,000 to over \$250,000 for a full-body tracking system.
Conventional motion capture datasets have involved expensive laboratory studies with up to 300 subjects paid to perform a variety of tasks while wearing a professional motion capture setup.

Motion capture data is also central to the operation of extended reality (XR) systems, which include devices supporting augmented reality (AR), virtual reality (VR), and mixed reality (MR) technologies. XR has experienced a recent surge in attention and popularity with the release of affordable self-contained VR devices like the Meta Quest series, as well as the recent announcement of the Apple Vision Pro. Most consumer-oriented virtual reality systems include a head-mounted display (HMD) and two hand-held controllers. The system uses either external or onboard sensors to measure the position and orientation of these devices in 3D space, providing six degrees of freedom (6DoF), captured at a rate of between 60 and 144 times per second.
In essence, XR devices have recently become an affordable and widely-adopted form of motion tracking system.

The motion data generated by an XR device is used by a client-side application, such as ``Beat Saber'' or ``Tilt Brush,'' to render auditory, visual, and haptic stimuli, creating an immersive 3D experience. In some cases, users capture and share recordings of the motion data generated during an XR usage session to allow other users to ``replay'' the same virtual experience.

\begin{figure}[h]
\centering
\begin{minipage}{.475\linewidth}
  \centering
  \includegraphics[width=\linewidth]{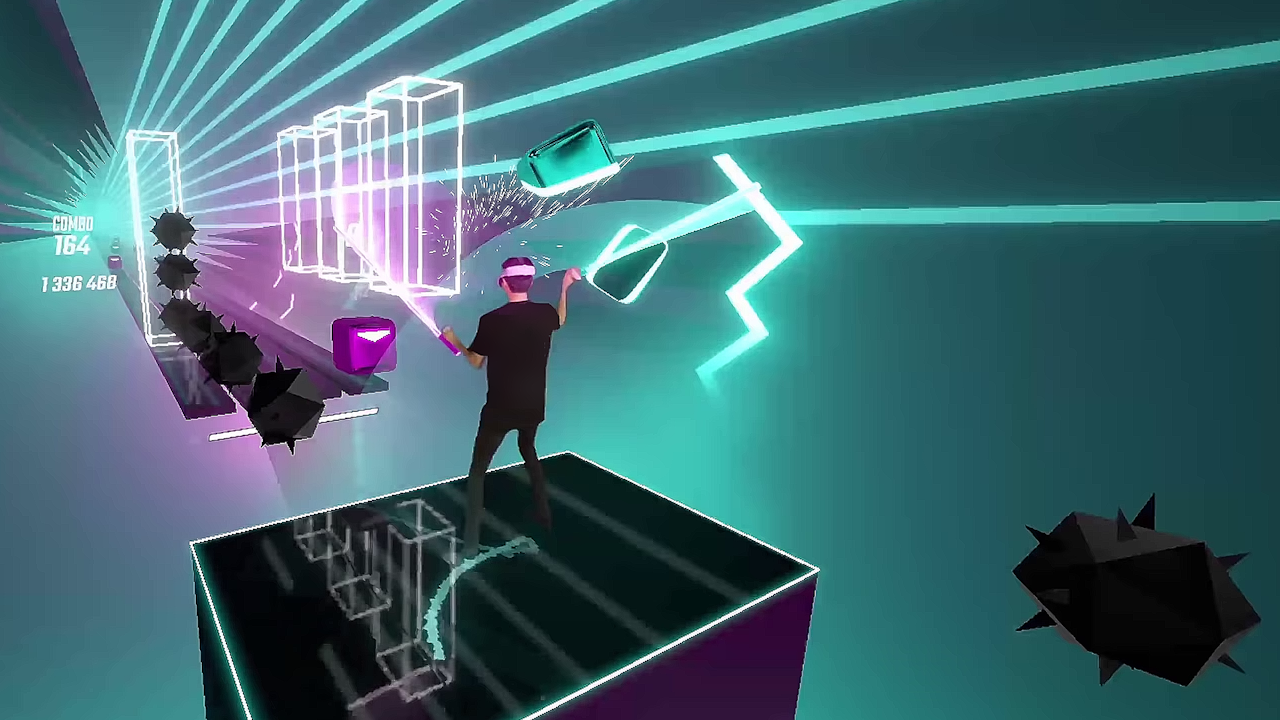}
  \captionof{figure}{``Beat Saber'' -- VR rhythm game.}
  \label{fig:beatsaber}
\end{minipage}%
\begin{minipage}{.05\linewidth}
    \includegraphics[width=0pt]{FIG-Beat-Saber.png}
\end{minipage}%
\begin{minipage}{.475\linewidth}
  \centering
  \includegraphics[width=\linewidth]{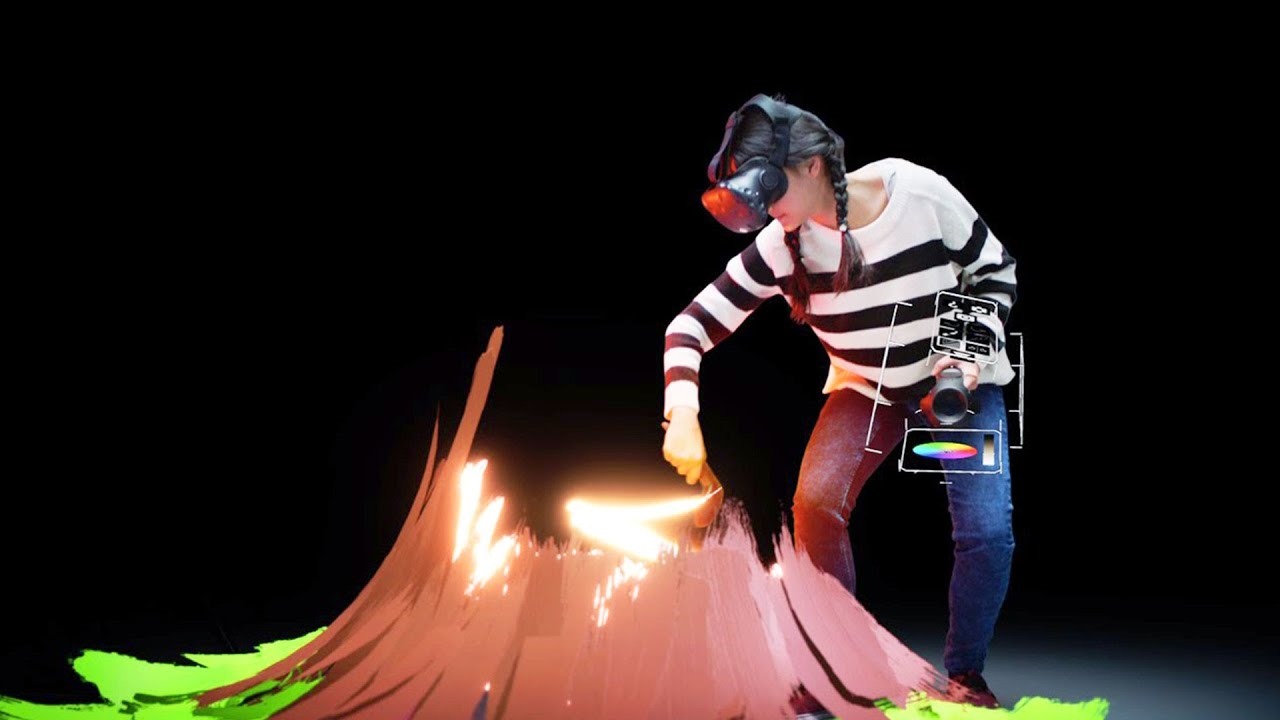}
  \captionof{figure}{``Tilt Brush'' -- VR painting app.}
  \label{fig:tiltbrush}
\end{minipage}
\end{figure}

\subsection{Beat Saber}
\vspace{-0.5em}
``Beat Saber'' \cite{beat_saber}, shown in Figure \ref{fig:beatsaber}, is a VR rhythm game where players slice blocks representing musical beats with a pair of sabers they hold in each hand. It is the primary data source for the BOXRR-23 dataset. With over 6 million copies sold, Beat Saber is the most popular VR application of all time \cite{wobbeking_beat_2022}.
The game contains a number of ``maps,'' which consist of an audio track and a series of objects presented to the user in time with the audio. These objects include ``blocks,'' which the player must hit at the correct angle with the correct saber, ``bombs,'' which the player must avoid hitting with their sabers, and ``walls,'' which the player must avoid with their head. The player is given a score based on their accuracy in completing these tasks. Reacting to these events typically requires users to deploy fast ballistic movements \cite{vitaladevuni_human_2007, douglas2012ergonomics}.

While hundreds of maps are included in the base game, over 100,000 user-created maps can be played by installing open-source game modifications. Beat Saber enthusiasts may choose to install open-source leaderboard extensions in order to compete with other players to achieve a higher ``rank'' on the leaderboards for popular maps. 
Two of the most popular Beat Saber leaderboard services are ``BeatLeader'' \cite{beatleader} and ``ScoreSaber'' \cite{scoresaber}, with a combined 4 million scores being submitted to the platforms to date. When submitting a score to either of these services, users attach a motion capture recording of them playing the corresponding Beat Saber map, which is then made publicly available on the BeatLeader or ScoreSaber website to allow others to audit the legitimacy of the claimed score.

\subsection{Tilt Brush}
\vspace{-0.5em}
``Tilt Brush'' \cite{tiltbrush}, shown in Figure \ref{fig:tiltbrush}, is a VR painting game created by Google that allows users to create 3D virtual objects using a variety of brushes and tools. Users can then export their drawings in various file formats, along with a motion capture recording of them creating the object, allowing other users to re-watch the original painting process. From 2017 to 2021, Google hosted ``Google Poly,'' a free service for sharing virtual creations (and accompanying motion capture recordings) from Tilt Brush. After the shutdown of Google Poly in 2021, the ``PolyGone'' project \cite{polygone} was created to host a free archive of over 50,000 user-submitted creations from Google Poly under a CC-BY license. Contrary to Beat Saber, Tilt Brush motion consists primarily of precise fine motor movements.

\section{Data Collection}
\label{collection}
\vspace{-0.5em}

\begin{figure}[h]
\centering
\includegraphics[width=0.75 \linewidth]{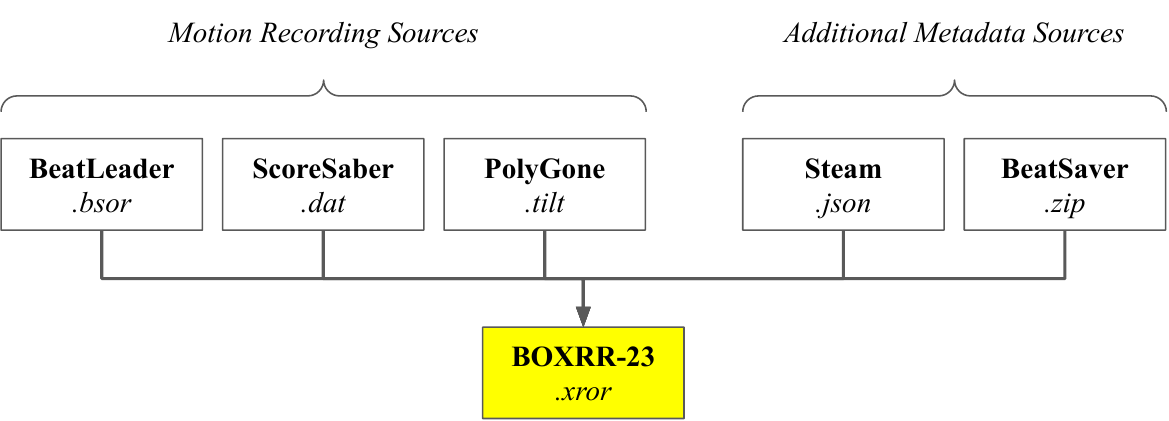}
\caption{Data collection and processing pipeline for BOXRR-23 dataset.}
\label{fig:pipeline}
\end{figure}
\vspace{-0.5em}

Figure \ref{fig:pipeline} shows the data collection process used to produce the BOXRR-23 dataset. We downloaded over 4.7 million publicly-available motion capture recordings stored on the BeatLeader, ScoreSaber, and PolyGone websites, and obtained additional metadata information, such as player experience levels and in-game events, from the public web APIs of Steam \cite{steam} and BeatSaver \cite{beatsaver}. We then removed identifiable details like player IDs and pseudonyms to protect the identity of each user.
Finally, we converted all recordings from their original formats into our purpose-built XROR format, described in \S\ref{xror}. The sizes of each of the sources, and of the dataset, are summarized in Table 1. We performed this data collection process in April 2023 and have included all valid, non-corrupt recordings submitted to all three platforms between November 1st, 2017 and April 15th, 2023.

\begin{table}[h]
\centering
\resizebox{0.75 \columnwidth}{!}{%
\begin{tabular}{llllll}
\multicolumn{6}{c}{Table 1(A): Sources for data in BOXRR-23 dataset.} \\ \hline
\multicolumn{1}{|l|}{\textbf{Source}} & \multicolumn{1}{l|}{\textbf{Application}} & \multicolumn{1}{l|}{\textbf{Users}} & \multicolumn{1}{l|}{\textbf{Recordings}} & \multicolumn{1}{l|}{\textbf{Format}} & \multicolumn{1}{l|}{\textbf{Size}} \\ \hline
\multicolumn{1}{|l|}{BeatLeader} & \multicolumn{1}{l|}{Beat Saber} & \multicolumn{1}{l|}{95,192} & \multicolumn{1}{l|}{3,525,456} & \multicolumn{1}{l|}{.bsor} & \multicolumn{1}{l|}{6.25 TB} \\ \hline
\multicolumn{1}{|l|}{ScoreSaber} & \multicolumn{1}{l|}{Beat Saber} & \multicolumn{1}{l|}{55,331} & \multicolumn{1}{l|}{1,136,581} & \multicolumn{1}{l|}{.dat} & \multicolumn{1}{l|}{1.44 TB} \\ \hline
\multicolumn{1}{|l|}{PolyGone} & \multicolumn{1}{l|}{Tilt Brush} & \multicolumn{1}{l|}{27,693} & \multicolumn{1}{l|}{55,178} & \multicolumn{1}{l|}{.tilt} & \multicolumn{1}{l|}{1.87 TB} \\ \hline
 &  &  &  &  &  \\
\multicolumn{6}{c}{Table 1(B): Output characteristics of BOXRR-23 dataset.} \\ \hline
\multicolumn{2}{|l|}{\textbf{Dataset}} & \multicolumn{1}{l|}{\textbf{Users}} & \multicolumn{1}{l|}{\textbf{Recordings}} & \multicolumn{1}{l|}{\textbf{Format}} & \multicolumn{1}{l|}{\textbf{Size}} \\ \hline
\rowcolor[HTML]{FFFF00} 
\multicolumn{2}{|l|}{\cellcolor[HTML]{FFFF00}BOXRR-23 Dataset} & \multicolumn{1}{l|}{\cellcolor[HTML]{FFFF00}105,852} & \multicolumn{1}{l|}{\cellcolor[HTML]{FFFF00}4,717,215} & \multicolumn{1}{l|}{\cellcolor[HTML]{FFFF00}.xror} & \multicolumn{1}{l|}{\cellcolor[HTML]{FFFF00}4.71 TB} \\ \hline
\end{tabular}%
}
\label{tab:sources}
\end{table}

\section{Related Work}

We searched for existing datasets relating to ``motion capture,'' ``telemetry,'' ``VR motion,'' ``XR motion,'' etc., on dataset hosting platforms like Kaggle, Zenodo, and Dryad, as well as for academic papers relating to motion capture data and experiments. We found over 25 existing datasets containing human motion recordings. The majority of these datasets come from conventional non-XR motion tracking systems, as listed in Table 2(A), while several originate from XR-based laboratory studies, listed in Table 2(B). The largest existing study contained 511 subjects \cite{TTI}, with a single session captured from each subject. By contrast, our dataset, summarized in Table 2(C), contains over 105,000 subjects and 4.7 million recordings from the three sources described in \S\ref{collection}.

In addition to being over 200 times larger than the largest existing dataset, we found that all of the existing datasets come from a laboratory study in which participants used a small number of homogeneous devices and were generally physically present in a narrow geographical area. Thus, the BOXRR-23 dataset is more useful for obtaining a representative sample of XR users, as it originates from real XR users using their own devices in their own homes. As a result, it contains diverse data from over 40 types of XR devices, and includes users from over 50 countries around the world.

\begin{table}[h]
\resizebox{\columnwidth}{!}{%
\begin{tabular}{llrrrr}
\multicolumn{5}{c}{Table 2(A): Current motion capture datasets outside XR.} \\ \hline
\multicolumn{1}{|l|}{\textbf{Dataset}} & \multicolumn{1}{l|}{\textbf{Organization}} & \multicolumn{1}{l|}{\textbf{Year}} & \multicolumn{1}{l|}{\textbf{Subjects}} & \multicolumn{1}{l|}{\textbf{Recordings}} & \multicolumn{1}{l|}{\textbf{Markers}} \\ \hline
\multicolumn{1}{|l|}{BMLrub \cite{Troje2002DecomposingBM}} & \multicolumn{1}{l|}{Ruhr University Bochum} & \multicolumn{1}{r|}{2002} & \multicolumn{1}{r|}{111} & \multicolumn{1}{r|}{3,061} & \multicolumn{1}{r|}{41, 3DoF} \\ \hline
\multicolumn{1}{|l|}{HDM05 \cite{inproceedings}} & \multicolumn{1}{l|}{Max Planck Society} & \multicolumn{1}{r|}{2007} & \multicolumn{1}{r|}{4} & \multicolumn{1}{r|}{215}& \multicolumn{1}{r|}{41, 3DoF} \\ \hline
\multicolumn{1}{|l|}{CMU-MMAC \cite{Torre2008GuideTT}} & \multicolumn{1}{l|}{Carnegie Mellon University} & \multicolumn{1}{r|}{2008} & \multicolumn{1}{r|}{5} & \multicolumn{1}{r|}{5} & \multicolumn{1}{r|}{41, 3DoF} \\ \hline
\multicolumn{1}{|l|}{EYES Japan \cite{mocapdata.com}} & \multicolumn{1}{l|}{EYES Japan} & \multicolumn{1}{r|}{2009} & \multicolumn{1}{r|}{12} & \multicolumn{1}{r|}{750} & \multicolumn{1}{r|}{37, 3DoF} \\ \hline
\multicolumn{1}{|l|}{HumanEva \cite{article}} & \multicolumn{1}{l|}{University of Toronto} & \multicolumn{1}{r|}{2010} & \multicolumn{1}{r|}{3} & \multicolumn{1}{r|}{28} & \multicolumn{1}{r|}{39, 3DoF} \\ \hline
\multicolumn{1}{|l|}{SFU MoCap \cite{mocap.cs.sfu.ca}} & \multicolumn{1}{l|}{Simon Fraser University} & \multicolumn{1}{r|}{2012} & \multicolumn{1}{r|}{7} & \multicolumn{1}{r|}{44} & \multicolumn{1}{r|}{53, 3DoF} \\ \hline
\multicolumn{1}{|l|}{ACCAD \cite{accad}} & \multicolumn{1}{l|}{Ohio State University} & \multicolumn{1}{r|}{2012} & \multicolumn{1}{r|}{20} & \multicolumn{1}{r|}{252} & \multicolumn{1}{r|}{82, 3DoF} \\ \hline
\multicolumn{1}{|l|}{Sleight of Hand \cite{10.1145/2159616.2159630}} & \multicolumn{1}{l|}{Trinity College Dublin} & \multicolumn{1}{r|}{2012} & \multicolumn{1}{r|}{1} & \multicolumn{1}{r|}{62} & \multicolumn{1}{r|}{91, 3DoF} \\ \hline
\multicolumn{1}{|l|}{Human3.6m \cite{6682899}} & \multicolumn{1}{l|}{Romanian Academy} & \multicolumn{1}{r|}{2013} & \multicolumn{1}{r|}{11} & \multicolumn{1}{r|}{44} & \multicolumn{1}{r|}{24, 3DoF} \\ \hline
\multicolumn{1}{|l|}{MoSh \cite{10.1145/2661229.2661273}} & \multicolumn{1}{l|}{Max Planck Society} & \multicolumn{1}{r|}{2014} & \multicolumn{1}{r|}{19} & \multicolumn{1}{r|}{77} & \multicolumn{1}{r|}{87, 3DoF} \\ \hline
\multicolumn{1}{|l|}{MPI Limits \cite{7298751}} & \multicolumn{1}{l|}{Max Planck Society} & \multicolumn{1}{r|}{2015} & \multicolumn{1}{r|}{3} & \multicolumn{1}{r|}{35} & \multicolumn{1}{r|}{53, 3DoF} \\ \hline
\multicolumn{1}{|l|}{KIT MoCap \cite{7251476}} & \multicolumn{1}{l|}{Karlsruhe Institute of Technology} & \multicolumn{1}{r|}{2016} & \multicolumn{1}{r|}{232} & \multicolumn{1}{r|}{2,925} & \multicolumn{1}{r|}{50, 3DoF} \\ \hline
\multicolumn{1}{|l|}{Total Capture \cite{Trumble:BMVC:2017}} & \multicolumn{1}{l|}{University of Surrey} & \multicolumn{1}{r|}{2017} & \multicolumn{1}{r|}{5} & \multicolumn{1}{r|}{37} & \multicolumn{1}{r|}{53, 3DoF} \\ \hline
\multicolumn{1}{|l|}{AMASS \cite{AMASS:ICCV:2019}} & \multicolumn{1}{l|}{Max Planck Society} & \multicolumn{1}{r|}{2019} & \multicolumn{1}{r|}{344} & \multicolumn{1}{r|}{11,265} & \multicolumn{1}{r|}{37, 3DoF} \\ \hline
\multicolumn{1}{|l|}{CMU MoCap \cite{cmu}} & \multicolumn{1}{l|}{Carnegie Mellon University} & \multicolumn{1}{r|}{2019} & \multicolumn{1}{r|}{144} & \multicolumn{1}{r|}{2,605} & \multicolumn{1}{r|}{41, 3DoF} \\ \hline
\multicolumn{1}{|l|}{MoVi \cite{Ghorbani_2021}} & \multicolumn{1}{l|}{Queen’s University} & \multicolumn{1}{r|}{2021} & \multicolumn{1}{r|}{90} & \multicolumn{1}{r|}{1,890} & \multicolumn{1}{r|}{12, 3DoF} \\ \hline
 &  & \multicolumn{1}{l}{} & \multicolumn{1}{l}{} & \multicolumn{1}{l}{} \\
\multicolumn{5}{c}{Table 2(B): Current motion capture datasets inside XR.} \\ \hline
\multicolumn{1}{|l|}{\textbf{Dataset}} & \multicolumn{1}{l|}{\textbf{Organization}} & \multicolumn{1}{l|}{\textbf{Year}} & \multicolumn{1}{l|}{\textbf{Subjects}} & \multicolumn{1}{l|}{\textbf{Recordings}} & \multicolumn{1}{l|}{\textbf{Trackers}} \\ \hline
\multicolumn{1}{|l|}{Behavioural Biometrics \cite{BehaviouralBiometrics}} & \multicolumn{1}{l|}{Bundeswehr University Munich} & \multicolumn{1}{r|}{2019} & \multicolumn{1}{r|}{22} & \multicolumn{1}{r|}{88} & \multicolumn{1}{r|}{3, 6DoF} \\ \hline
\multicolumn{1}{|l|}{TTI \cite{TTI}} & \multicolumn{1}{l|}{Stanford University} & \multicolumn{1}{r|}{2020} & \multicolumn{1}{r|}{511} & \multicolumn{1}{r|}{511} & \multicolumn{1}{r|}{3, 6DoF} \\ \hline
\multicolumn{1}{|l|}{Body Normalization \cite{normal}} & \multicolumn{1}{l|}{University of Duisburg-Essen} & \multicolumn{1}{r|}{2021} & \multicolumn{1}{r|}{16} & \multicolumn{1}{r|}{48} & \multicolumn{1}{r|}{3, 6DoF} \\ \hline
\multicolumn{1}{|l|}{Obfuscation \cite{9583839}} & \multicolumn{1}{l|}{University of Central Florida} & \multicolumn{1}{r|}{2021} & \multicolumn{1}{r|}{60} & \multicolumn{1}{r|}{120} & \multicolumn{1}{r|}{3, 6DoF} \\ \hline
\multicolumn{1}{|l|}{Body Sway \cite{sway}} & \multicolumn{1}{l|}{Purdue University} & \multicolumn{1}{r|}{2021} & \multicolumn{1}{r|}{28} & \multicolumn{1}{r|}{336} & \multicolumn{1}{r|}{3, 6DoF} \\ \hline
\multicolumn{1}{|l|}{You Can’t Hide \cite{tricomi2022cant}} & \multicolumn{1}{l|}{University of Padova} & \multicolumn{1}{r|}{2022} & \multicolumn{1}{r|}{35} & \multicolumn{1}{r|}{69} & \multicolumn{1}{r|}{3, 6DoF} \\ \hline
\multicolumn{1}{|l|}{Motion Matching \cite{ponton2022mmvr}} & \multicolumn{1}{l|}{Technical University of Catalonia} & \multicolumn{1}{r|}{2022} & \multicolumn{1}{r|}{1} & \multicolumn{1}{r|}{12} & \multicolumn{1}{r|}{3, 6DoF} \\ \hline
\multicolumn{1}{|l|}{Personal Identifiability \cite{miller2023largescale}} & \multicolumn{1}{l|}{Stanford University} & \multicolumn{1}{r|}{2023} & \multicolumn{1}{r|}{232} & \multicolumn{1}{r|}{1856} & \multicolumn{1}{r|}{3, 6DoF} \\ \hline
\multicolumn{1}{|l|}{Who is Alyx \cite{christian_schell_2023_7663984}} & \multicolumn{1}{l|}{University of Würzburg} & \multicolumn{1}{r|}{2023} & \multicolumn{1}{r|}{71} & \multicolumn{1}{r|}{142} & \multicolumn{1}{r|}{3, 6DoF} \\ \hline
 &  & \multicolumn{1}{l}{} & \multicolumn{1}{l}{} & \multicolumn{1}{l}{} \\
\multicolumn{5}{c}{Table 2(C): Our new XR motion capture dataset.} \\ \hline
\multicolumn{1}{|l|}{\textbf{Dataset}} & \multicolumn{1}{l|}{\textbf{Organization}} & \multicolumn{1}{l|}{\textbf{Year}} & \multicolumn{1}{l|}{\textbf{Subjects}} & \multicolumn{1}{l|}{\textbf{Recordings}} & \multicolumn{1}{l|}{\textbf{Trackers}} \\ \hline
\rowcolor[HTML]{FFFF00} 
\multicolumn{1}{|l|}{\cellcolor[HTML]{FFFF00}BOXRR-23} & \multicolumn{1}{l|}{\cellcolor[HTML]{FFFF00}University of California, Berkeley} & \multicolumn{1}{r|}{\cellcolor[HTML]{FFFF00}2023} & \multicolumn{1}{r|}{\cellcolor[HTML]{FFFF00}105,852} & \multicolumn{1}{r|}{\cellcolor[HTML]{FFFF00}4,717,215} & \multicolumn{1}{r|}{\cellcolor[HTML]{FFFF00}3, 6DoF} \\ \hline
\end{tabular}%
}
\label{tab:mocap}
\end{table}

As evidenced by Table 2, BOXRR-23 is more comparable to existing XR datasets with a small number of 6DoF trackers than non-XR datasets with a large number of 3DoF markers. In applications where detailed full-body tracking is required, a conventional MoCap dataset may be more appropriate.

\eject

\section{XROR Format}
\label{xror}

As detailed in \S\ref{collection}, the data included in the BOXRR-23 dataset was scraped from three separate sources (BeatLeader, ScoreSaber, and PolyGone), each using three separate custom file formats designed specifically for those platforms (.BSOR, .DAT, and .TILT, respectively, summarized in Table 3(A)). We felt that the experience of future consumers of this dataset would be improved if the recordings were all converted to a single file format that could be analyzed and ingested via a unified pipeline. 

We began by evaluating open-source motion capture file formats such as .BVA, .BVH, and .MVNX. Unfortunately, we found that the existing formats were unsuitable for this database for a variety of reasons. Some formats, such as .BVA and .BVH, only have support for motion data, and did not allow us to embed the rich metadata and event data streams we wished to include in the dataset. Others, like .MVNX, did support the inclusion of arbitrary metadata and event data streams, but used an inefficient underlying text-based file format (.XML) that would have caused the dataset to balloon to over 300 TB in size. Finally, some proprietary formats did contain all of the necessary features in an efficient binary format, but were not open-source and required paid tools or licenses to utilize them. Overall, we found that none of the existing open-source file formats were unsuitable for this dataset.

A formal specification of the XROR format, using the BSON version of the JSON Schema notation, is provided here: \url{https://rdi.berkeley.edu/metaverse/boxrr-23/dict.json}.

To address the issues with existing open-source file formats, we introduce the new ``Extened Reality Open Recording (XROR)'' file format. XROR files contain metadata as well as rich event and motion data streams, and are based internally on BSON (Binary JSON), a flexible, widely-supported format with libraries in dozens of languages. Metadata is stored as JSON key-value pairs, while event data and motion data streams are converted to 2D floating-point arrays and compressed using fpzip, a lossless compressor of multidimensional floating-point arrays designed by Lawrence Livermore National Laboratory specifically for the efficient storage and transmission of scientific datasets.

To evaluate the relative efficiency of our new format, we converted a portion of our dataset into a variety of existing open formats, summarized in Table 3(B), as well as our proposed XROR format, as shown in Table 3(C). Even compared to the original, purpose-built formats shown in Table 3(A), XROR achieves space savings of at least 30\% with no loss in precision.

\begin{table}[h]
\centering
\resizebox{0.75 \columnwidth}{!}{%
\begin{tabular}{lccccl}
\multicolumn{6}{c}{Table 3(A): Source file formats for motion data.} \\ \hline
\multicolumn{1}{|l|}{\textbf{Format}} & \multicolumn{1}{c|}{\textbf{Metadata}} & \multicolumn{1}{c|}{\textbf{Motion Data}} & \multicolumn{1}{c|}{\textbf{Event Data}} & \multicolumn{1}{c|}{\textbf{Compression}} & \multicolumn{1}{l|}{\textbf{Avg. Size}} \\ \hline
\multicolumn{1}{|l|}{.tilt} & \multicolumn{1}{c|}{\Checkmark} & \multicolumn{1}{c|}{\Checkmark} & \multicolumn{1}{c|}{\Checkmark} & \multicolumn{1}{c|}{} & \multicolumn{1}{l|}{33.89 MB} \\ \hline
\multicolumn{1}{|l|}{.bsor} & \multicolumn{1}{c|}{\Checkmark} & \multicolumn{1}{c|}{\Checkmark} & \multicolumn{1}{c|}{\Checkmark} & \multicolumn{1}{c|}{} & \multicolumn{1}{l|}{1.77 MB} \\ \hline
\multicolumn{1}{|l|}{.dat} & \multicolumn{1}{c|}{\Checkmark} & \multicolumn{1}{c|}{\Checkmark} & \multicolumn{1}{c|}{\Checkmark} & \multicolumn{1}{c|}{} & \multicolumn{1}{l|}{1.27 MB} \\ \hline
 & \multicolumn{1}{l}{} & \multicolumn{1}{l}{} & \multicolumn{1}{l}{} & \multicolumn{1}{l}{} &  \\
\multicolumn{6}{c}{Table 3(B): Existing general file formats for motion data.} \\ \hline
\multicolumn{1}{|l|}{\textbf{Format}} & \multicolumn{1}{c|}{\textbf{Metadata}} & \multicolumn{1}{c|}{\textbf{Motion Data}} & \multicolumn{1}{c|}{\textbf{Event Data}} & \multicolumn{1}{c|}{\textbf{Compression}} & \multicolumn{1}{l|}{\textbf{Avg. Size}} \\ \hline
\multicolumn{1}{|l|}{.mvnx} & \multicolumn{1}{c|}{\Checkmark} & \multicolumn{1}{c|}{\Checkmark} & \multicolumn{1}{c|}{\Checkmark} & \multicolumn{1}{c|}{} & \multicolumn{1}{l|}{61.90 MB} \\ \hline
\multicolumn{1}{|l|}{.bvh} & \multicolumn{1}{c|}{} & \multicolumn{1}{c|}{\Checkmark} & \multicolumn{1}{c|}{} & \multicolumn{1}{c|}{} & \multicolumn{1}{l|}{25.79 MB} \\ \hline
\multicolumn{1}{|l|}{.bva} & \multicolumn{1}{c|}{} & \multicolumn{1}{c|}{\Checkmark} & \multicolumn{1}{c|}{} & \multicolumn{1}{c|}{} & \multicolumn{1}{l|}{13.98 MB} \\ \hline
 &  &  &  &  &  \\
\multicolumn{6}{c}{Table 3(C): Proposed new open file format for motion data.} \\ \hline
\multicolumn{1}{|l|}{\textbf{Format}} & \multicolumn{1}{c|}{\textbf{Metadata}} & \multicolumn{1}{c|}{\textbf{Motion Data}} & \multicolumn{1}{c|}{\textbf{Event Data}} & \multicolumn{1}{c|}{\textbf{Compression}} & \multicolumn{1}{l|}{\textbf{Avg. Size}} \\ \hline
\rowcolor[HTML]{FFFF00} 
\multicolumn{1}{|l|}{\cellcolor[HTML]{FFFF00}.xror} & \multicolumn{1}{c|}{\cellcolor[HTML]{FFFF00}\Checkmark} & \multicolumn{1}{c|}{\cellcolor[HTML]{FFFF00}\Checkmark} & \multicolumn{1}{c|}{\cellcolor[HTML]{FFFF00}\Checkmark} & \multicolumn{1}{c|}{\cellcolor[HTML]{FFFF00}\Checkmark} & \multicolumn{1}{l|}{\cellcolor[HTML]{FFFF00}0.99 MB} \\ \hline
\end{tabular}%
}
\label{tab:formats}
\end{table}

Due to the advantages of our new XROR format over the existing alternatives, the entire BOXRR-23 dataset is offered exclusively as XROR files. To help researchers process this format, we have provided open-source tools to parse XROR files, and convert them to and from a variety of formats, including .TILT, .BSOR, .DAT, and .JSON: \url{https://github.com/metaguard/xror}.

\eject

\section{Recording Contents}

\vspace{-1em}

\begin{figure}[h]
\centering
\begin{minipage}{.475\linewidth}
  \centering
  \includegraphics[height=15em]{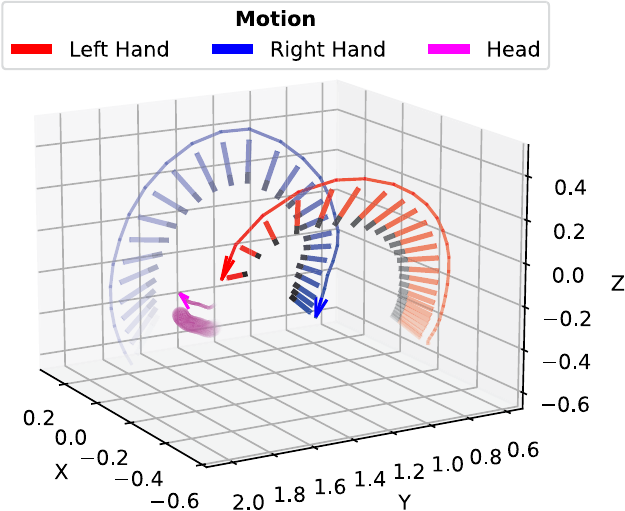}
  \captionof{figure}{``Beat Saber'' motion data.}
  \label{fig:beatsaber_motion}
\end{minipage}%
\begin{minipage}{.05\linewidth}
    \includegraphics[width=0pt]{FIG-Beat-Saber.png}
\end{minipage}%
\begin{minipage}{.475\linewidth}
  \centering
  \includegraphics[height=15em]{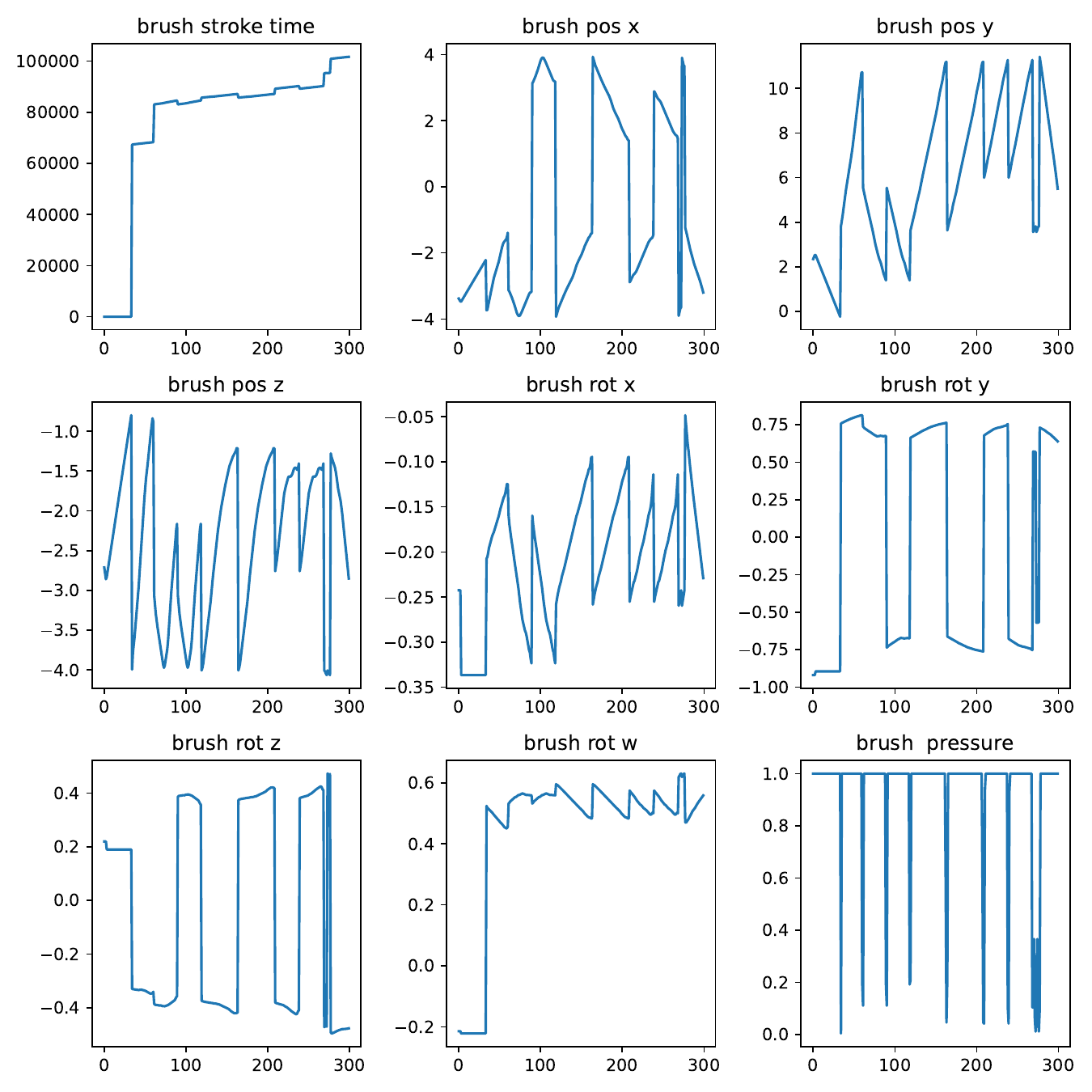}
  \captionof{figure}{``Tilt Brush'' motion data.}
  \label{fig:tiltbrush_motion}
\end{minipage}
\end{figure}

\vspace{-1em}

\begin{figure}[h]
\centering
\begin{minipage}{.475\linewidth}
  \centering
  \includegraphics[height=15em]{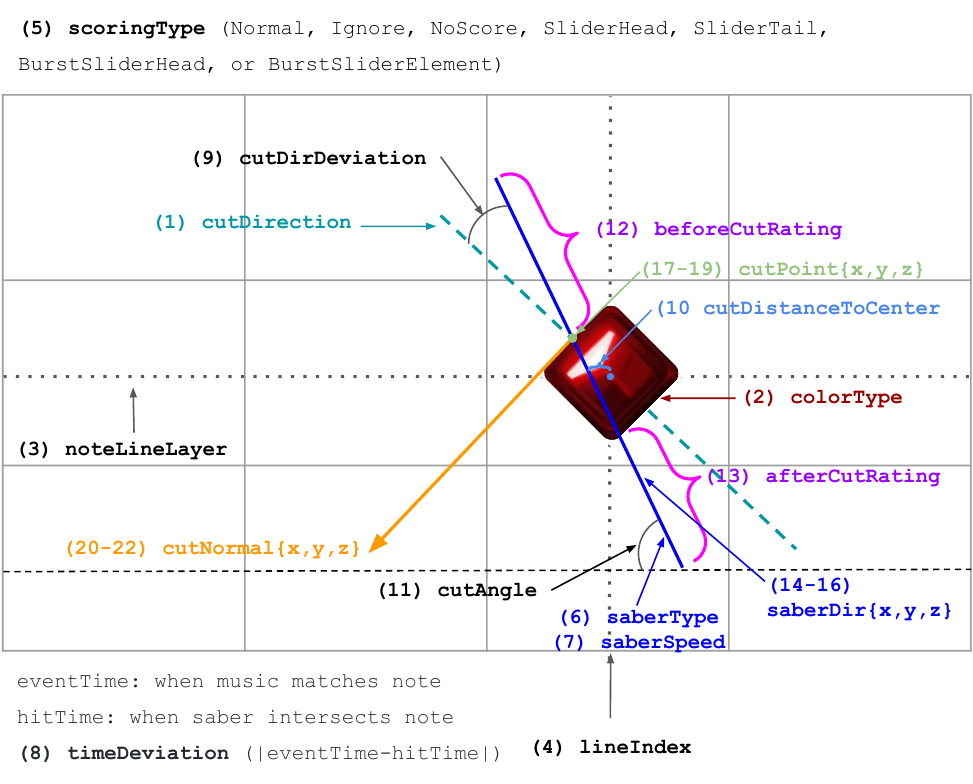}
  \captionof{figure}{``Beat Saber'' event data.}
  \label{fig:beatsaber_events}
\end{minipage}%
\begin{minipage}{.05\linewidth}
    \includegraphics[width=0pt]{FIG-Beat-Saber.png}
\end{minipage}%
\begin{minipage}{.475\linewidth}
  \centering
  \includegraphics[height=15em]{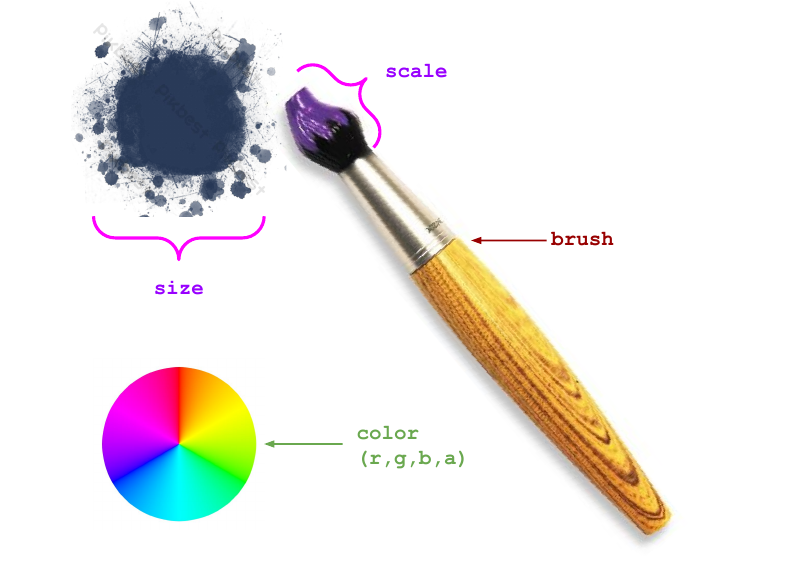}
  \captionof{figure}{``Tilt Brush'' event data.}
  \label{fig:tiltbrush_events}
\end{minipage}
\end{figure}

Figures \ref{fig:beatsaber_motion}--\ref{fig:tiltbrush_events} illustrate the typical contents of each recording in the BOXRR-23 dataset. Specifically, the following data is included in each recording:

\begin{enumerate}[leftmargin=*]
    \itemsep 0em
    \item \textbf{Metadata}. A variety of metadata is included with each entry, including anonymized user IDs, hardware and software descriptions, and virtual environment and activity descriptions.
    \item \textbf{Motion data}. Recordings principally consist of motion data captured in 6DoF at between 60~Hz and 144~Hz. Beat Saber recordings include head and hand motion data (see Fig. \ref{fig:beatsaber_motion}), while Tilt Brush recordings include brush motion and pressure data (see Fig. \ref{fig:tiltbrush_motion}).
    \item \textbf{Event data}. Motion data is accompanied by rich contextual information about events occurring in the virtual world. This includes information about the in-game objects and obstacles in the case of Beat Saber (see Fig. \ref{fig:beatsaber_events}), and about each brush stroke in the case of Tilt Brush (see Fig. \ref{fig:tiltbrush_events}).
\end{enumerate}

Data examples of Beat Saber and Tilt Brush recordings are provided in the supplemental materials.

\section{Access Instructions}
\label{sec:access}

Researchers interested in using the BOXRR-23 dataset are invited to visit \url{https://rdi.berkeley.edu/metaverse/boxrr-23/}. The permanent DOI is \url{https://doi.org/10.25350/B5NP4V}.
For ease of access, the dataset has been split into 106 .zip files, each containing up to 1,000 users. Each user is represented by a folder, containing one or more recordings from that user in .xror format.

We developed the licensing terms for this dataset in conjunction with the Committee for Protection of Human Subjects (CPHS) and Intellectual Property \& Industry Research Alliances (IPIRA) groups at UC Berkeley, with the chief goal of protecting the human subjects contained in this dataset. The dataset is licensed under a Creative Commons Attribution-NonCommercial-ShareAlike 4.0 International (CC BY-NC-SA 4.0) license, and is additionally subject to an ethical data use agreement (DUA) that prohibits unethical uses of the data, such as attempts to deanonymize the subjects. Access to the dataset is automatically granted upon agreeing to the CC BY-NC-SA 4.0 license and DUA.

\section{Intended Use Cases}
As detailed in \S\ref{intro}, we originally produced this dataset for use in a VR authentication study, which required a large number of users for comparison with traditional biometrics. However, there are a number of interesting uses for this dataset beyond security and privacy research.

\subsection{Notable Known Uses}
\label{known_uses}

Until recently, this dataset has only been available for internal use at UC Berkeley. Thus far, we have published three papers using this dataset in the XR security and privacy domain:

\begin{itemize}[leftmargin=*]
    \item We conducted a study that uniquely identified over 55,000 VR users based on their head and hand motion \cite{nair2023unique}. By using the BOXRR-23 dataset, this study was over 200 times larger than the next largest VR identification study, and the first to demonstrate parity with biometrics like fingerprints.
    \begin{itemize}[leftmargin=*]
        \item Result: After training a classification model on 5 minutes of data per person, a user can be uniquely identified amongst the entire pool with 94.33\% accuracy from 100 seconds of motion.
        \item Availability: The source code and documentation required to replicate this result using the BOXRR-23 dataset can be found at \url{https://github.com/metaguard/identification}. 
    \end{itemize}
    \item In another study, we combined the BOXRR-23 dataset with a survey to demonstrate that a large number of sensitive data attributes can be inferred from VR users based on motion alone \cite{nair2023inferring}.
    \begin{itemize}[leftmargin=*]
        \item Result: Using simple machine learning models, over $35$ private data attributes could accurately and consistently be inferred from VR users using head and hand motion data alone.
    \end{itemize}
    \item In a third paper, we presented ``MetaGuard,'' \cite{nair2023going} a differential privacy-based tool for protecting user data privacy in the metaverse, which we evaluated using the BOXRR-23 dataset.
    \begin{itemize}[leftmargin=*]
        \item Result: We show a significant degradation of attacker capabilities when using MetaGuard.
        \item Availability: The source code and documentation required to replicate this result using the BOXRR-23 dataset can be found at \url{https://github.com/metaguard/metaguard}. 
    \end{itemize}
\end{itemize}

\subsection{Future Directions}
While the dataset was originally intended for use in the security and privacy domain, and has thus far only been used in this field, we can envision a number of additional interesting applications for this data. Historically, motion capture data has primarily been used for computer graphics, animation, and CGI, and our data could also be used in this domain. For example, it could be used to train large-scale generative machine learning models for natural human motion synthesis tasks. It may also be of interest to researchers studying human-computer interaction in XR. For example, researchers could use the data to investigate interaction patterns likely to cause discomfort or injury.

One area of active research that is relevant to our dataset is the inference of full-body pose information from sparse tracking inputs. Researchers have demonstrated the ability to recover full-body motion data from the motion of a few tracked points \cite{jiang2022avatarposer, du2023avatars}. Using these techniques, the sparse tracking data offered by our dataset could be used to recover inferred full-body motion for a variety of applications.

Furthermore, the dataset contains numerous labels, including anonymized user IDs, hardware and software descriptions, and virtual environment and activity descriptions, that can be used to construct novel classification and regression tasks. For example, a very interesting use of the Tilt Brush portion of the dataset could be to use the brushstroke motion data to infer the title or description of the drawing, which are provided in the metadata as potential labels.

Finally, this dataset presents a challenging and unique opportunity for theoretical machine learning research, because it consists of long, sequential data, with sequence lengths often in excess of 100,000. Most existing deep learning algorithms are not well equipped to handle sequential data of this size. Currently, our dataset is a rare instance of a task in which classical ML algorithms seem to outperform deep learning methods \cite{nair2023unique}. Developing models that can accurately and efficiently ingest the data contained in this dataset may require theoretical advances in machine learning techniques.



\section{Population Survey}
\label{survey}

To shed additional light on the demographics of the users within our dataset, we conducted a large-scale online survey of VR users. The survey contained about 50 questions and received 1,006 responses, of which 830 users were present in the BOXRR-23 dataset. It was conducted in coordination with BeatLeader and other Beat Saber organizations, and thus did not reach the 1\% of BOXRR-23 users from Tilt Brush. The full results of this survey are available at \url{https://arxiv.org/abs/2305.14320}, and are summarized in Figure \ref{fig:survey} below.

\begin{figure}[h]
\centering
\begin{minipage}{.33\linewidth}
  \centering
  \includegraphics[width=\linewidth, trim=1.25in 0in 1in 0in, clip]{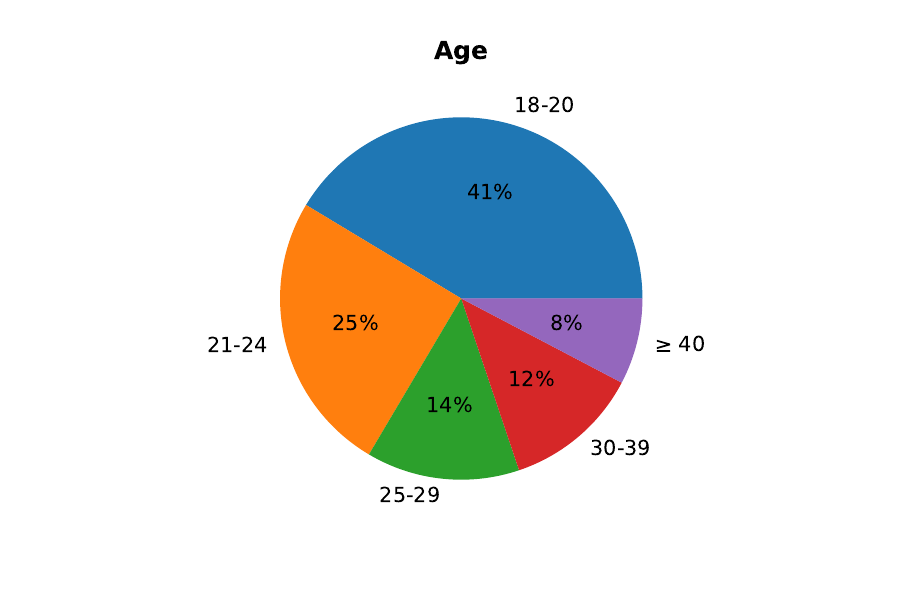}
  \includegraphics[width=\linewidth, trim=1.25in 0in 1in 0in, clip]{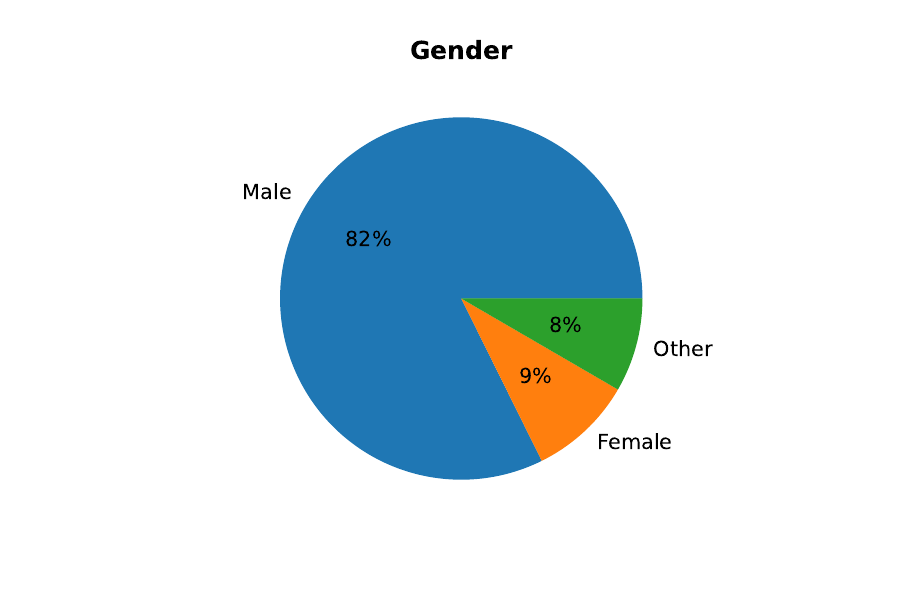}
\end{minipage}%
\begin{minipage}{.33\linewidth}
  \centering
  \includegraphics[width=\linewidth, trim=1.25in 0in 1in 0in, clip]{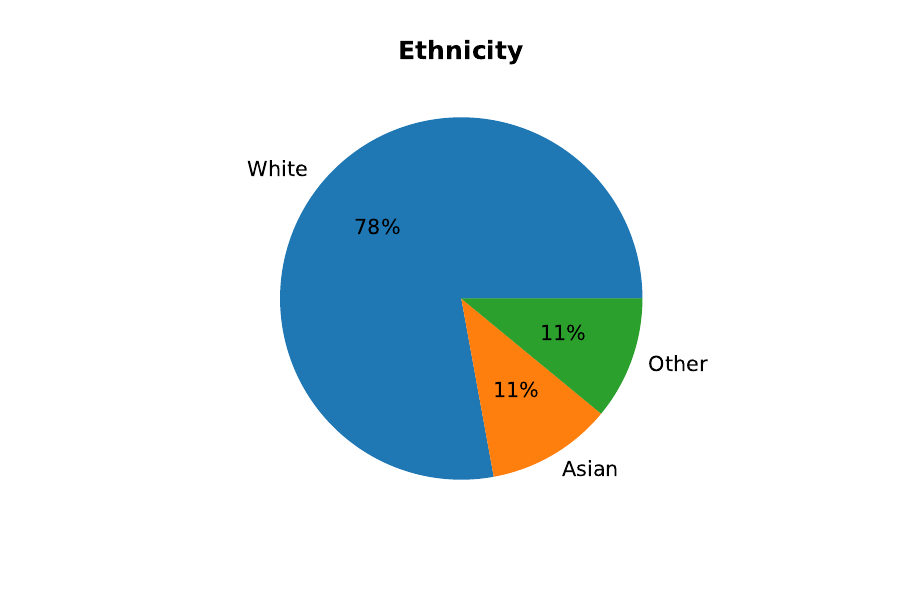}
  \includegraphics[width=\linewidth, trim=1.25in 0in 1in 0in, clip]{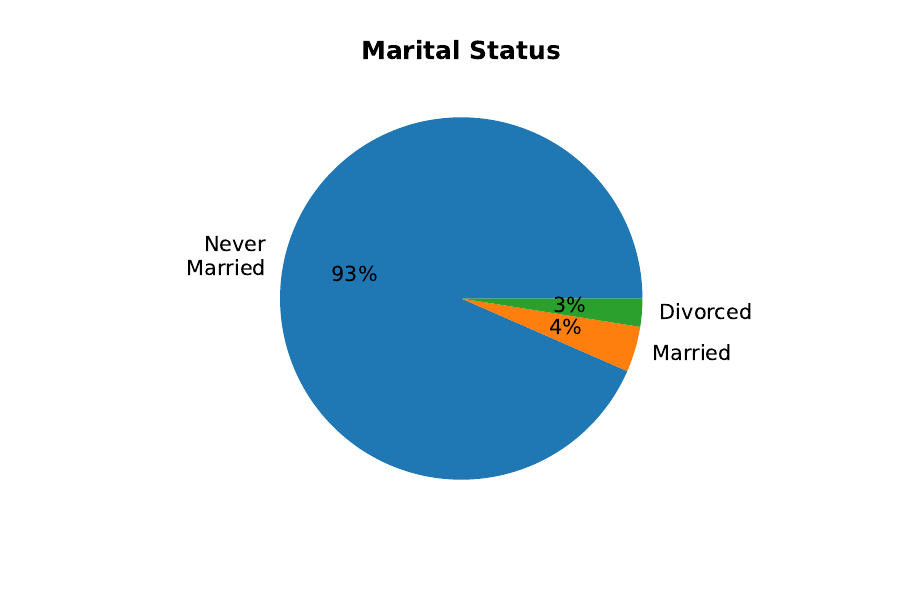}
\end{minipage}
\begin{minipage}{.33\linewidth}
  \centering
  \includegraphics[width=\linewidth, trim=1.25in 0in 1in 0in, clip]{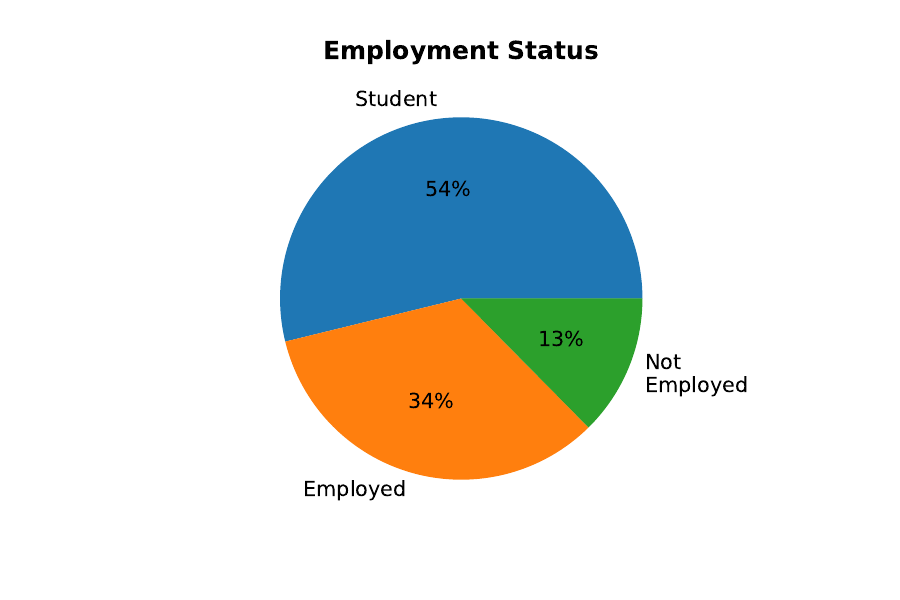}
  \includegraphics[width=\linewidth, trim=1.25in 0in 1in 0in, clip]{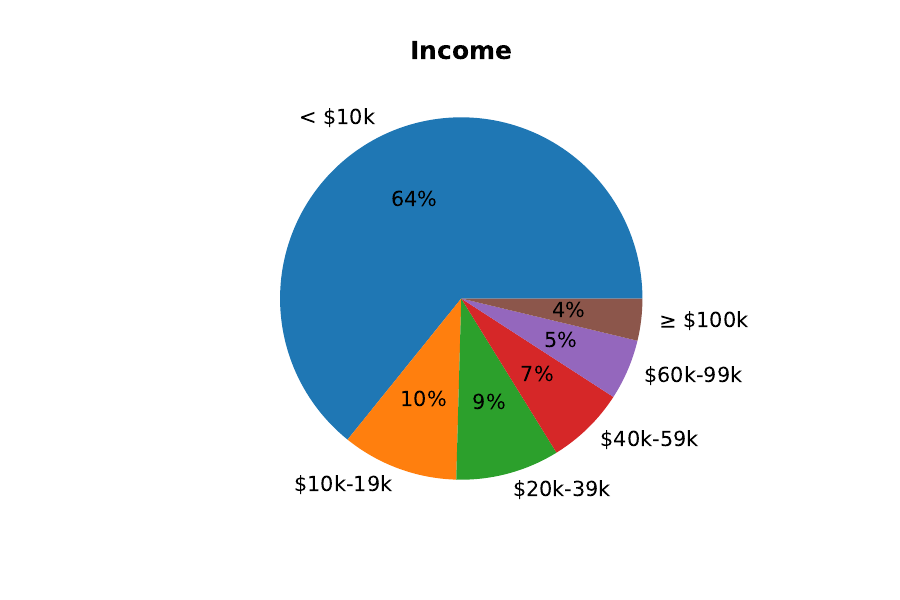}
\end{minipage}
\caption{Survey results from 830 users present in the BOXRR-23 dataset.}
\label{fig:survey}
\end{figure}

\section{Limitations}
\label{limitations}

As may be evident by the survey results provided in \S\ref{survey}, the users included in our dataset are not necessarily representative of a general population. For example, the dataset consists primarily of white and male subjects.
While the subjects are demographically similar to the overall population of VR device users \cite{noauthor_report_2017}, they consist entirely of users who chose to upload a BeatSaber performance or TiltBrush drawing to a public platform. As such, we believe enthusiast or expert-level users are likely to be overrepresented in the dataset. However, for the same reason, the dataset likely contains far more geographic diversity than existing laboratory-based datasets.
Furthermore, the data is derived from just two VR applications, Beat Saber and Tilt Brush, with almost 75\% of the users and 99\% of the recordings being from Beat Saber alone.
Overall, researchers should be cautious when attempting to use this dataset to draw conclusions about larger populations than the ones directly included. When attempting to use BOXRR-23 to draw conclusions about broader populations, researchers are advised to follow known best practices for accounting for sampling bias in machine learning datasets \cite{pagano2022bias, koch2021reduced}.

\eject

Additionally, there are some risks associated with the dataset being derived from ordinary XR users. Some metadata values, such as Beat Saber song titles or Tilt Brush drawing descriptions, may contain objectionable content due to their user-submitted nature. Metadata constituting user-configured settings like height and handedness should be considered self-reported, and are subject to the typical response biases associated with self-reported values. Finally, because the data is from ``the wild'' rather than a laboratory study, it originates from a wide variety of heterogeneous XR devices and physical environments, and may include more noise and tracking errors than a lab-created dataset.

\section{Ethical Considerations}
\label{ethics}

Because our dataset consists entirely of motion capture recordings from human subjects, significant attention was given to ethics throughout the process of designing and collecting the dataset.
Our collection of this dataset was overseen by the UC Berkeley Office for Protection of Human Subjects (OPHS), an OHRP-certified Institutional Review Board (IRB), approved as protocol \#2023-03-16120.

We note that in producing this dataset, the authors had no direct contact with human subjects. Instead, our data is derived from three public sources. All data utilized in this study was already broadly, publicly available, to any person in the world with an internet connection, without the need for permissions, credentials, authentication, or any special tools or applications, via the websites of ScoreSaber, BeatLeader, and PolyGone. No new data is being made accessible to the public in the publication of this dataset; our contribution is in finding, scraping, aggregating, reprocessing, enriching, and distributing this existing data, and in surveying the underlying population.

Despite the public nature of the data and the IRB approval, we chose to obtain written permission from ScoreSaber, BeatLeader, and PolyGone before proceeding out of an abundance of caution and respect for the communities from which this data originates. We did not begin collecting data until authorized to do so by these communities, and sought their input throughout the collection process.

Users of the ScoreSaber, BeatLeader, and PolyGone platforms must voluntary install custom software to share their motion recording data with these platforms. They are fully aware of the nature of the data being shared, as uploading and publicly sharing XR data is the explicit purpose of these platforms. They also consent to their recordings being made publicly available in the privacy policies of these platforms. For example, the BeatLeader Privacy Policy, which can be found at \url{https://www.beatleader.xyz/privacy}, states that ``Replays may contain personally identifiable information... Your data, including associated personally identifiable information, will be broadly publicly available to anyone with an internet connection via the BeatLeader website.'' Users of Google Poly (and PolyGone) consent to making their data publicly available under a CC-BY license.

Beyond consenting to the publication of their data in privacy policies and license agreements, we made further attempts to notify users of their involvement in academic research. Because users authenticate with these platforms via OAuth, their contact information is not known to the platforms, making direct consultation infeasible. However, we worked in collaboration with the BeatLeader team to inform users of their inclusion in academic research via their website and the official social media channels of the platform, and to develop an opt-out mechanism.

Although users knowingly consented to the public availability of their motion data, we took two additional steps to protect the privacy of data subjects. First, all known explicit identifiers, such as usernames and user IDs, have been removed from the dataset. No potentially sensitive information, such as protected health information, is included in the data or metadata. Second, the dataset is offered under a data use agreement (DUA) that prohibits researchers from attempting to deanonymize or contact the users, or to infer private attributes of the users that may be deemed sensitive.
We voluntarily followed the strictest PII data handling standards and guidelines offered by our institution throughout the dataset collection process to preclude the accidental release of non-anonymized data.

\eject

Participants originally submitted their motion data to the ScoreSaber, BeatLeader, and PolyGone platforms for purposes other than academic research. Namely, they chose to make their data freely publicly available for reasons such as competitive e-sports or collaborative artwork; as such, users were not compensated for their original submissions, nor for their inclusion in the dataset. Moreover, any participant risks associated with the use of an extended reality device would have been realized by the users regardless of the later inclusion of the resultant motion recordings in this dataset.
The scraping and redistribution of publicly-available online data is a highly common and widely accepted practice within the machine learning community \cite{rahman2021an, fan2022minedojo}.

While it is impossible to entirely eliminate the risks associated with a new dataset, we believe the additional risk posed by our dataset is minimal in light of the fact that all of the included data was already public. On the other hand, the data has the potential to facilitate significant advances in fields like graphics, HCI, XR, AI/ML, and computer security and privacy.
We have taken significant steps to mitigate the potential harms of this dataset while maximizing its utility for beneficial research.
Overall, we believe this research constitutes a net benefit to the subjects whose data was included by shedding light on the implications of the motion capture data which they have already, independently chosen to publish.
For instance, security and privacy research using this dataset benefits society by highlighting the magnitude of the VR privacy threat and motivating future work on countermeasures.

\vspace{-0.5em}

\section{Conclusion}
\vspace{-0.5em}

We have presented the BOXRR-23 dataset, a 4.7 TB dataset of extended reality motion capture recordings from users around the world.
Unlike existing motion capture datasets, BOXRR-23 is derived from recordings submitted by participants using their own XR devices, rather than a laboratory setup. As a result, it contains over 200 times more users, and over 400 times more recordings, than all known comparable datasets, while simultaneously being more diverse and ecologically valid.

The two XR applications included in BOXRR-23, Beat Saber and Tilt Brush, provide highly complementary motion data. Beat Saber consists almost entirely of fast ballistic movements while Tilt Brush consists almost entirely of fine motor movements, each controlled by a separate part of the brain \cite{fromm1977relation}.
By combining these sources, BOXRR-23 provides researchers a diverse collection of motion patterns.

For the first time, BOXRR-23 allows the identifiability of human motion data to be directly compared with biometrics like fingerprints and facial recognition, which have long enjoyed large public datasets. As such, we hope to see new advances in passive authentication mechanisms and privacy-preserving systems for XR, in addition to potential deployments in fields ranging from graphics and animation to usability and human-computer interaction.

In addition to identifying three new sources of motion data not previously widely known to academic researchers, we contributed a new XROR format to enable the efficient storage and transmission of this data. Our XROR format is approximately 30\% more efficient than the three original data formats, without any loss in precision, while also being more versatile than most existing open-source formats.
Documentation for our dataset is offered according to widely-recognized open standards, including Datasheets for Datasets \cite{gebruDatasheetsDatasets2020} and Dataset Nutrition Labels \cite{holland2018dataset}.
We also conducted a large survey of over 800 users present in the dataset to help researchers understand its demographic constituency.

As advances in extended reality allow this technology to reach increasingly large audiences, human motion data will remain vital to the operation XR and ``metaverse'' systems for the foreseeable future. In particular, augmented reality (AR) technology promises to be the next major medium of human-computer interactions, potentially even replacing the use of mobile devices such as smartphones. If this reality comes to pass, it is vital that we improve our understanding of the uses and implications of the motion data that these devices are designed to generate. We look forward to seeing future work that deploys our dataset to advance public knowledge in a variety of important fields, and to drive improvements to XR and metaverse experiences that benefit the field of extended reality as a whole.

\eject

\begin{ack}
This work was supported in part by the National Science Foundation (NSF), the National Physical Science Consortium (NPSC), the Fannie and John Hertz Foundation, and the Berkeley Center for Responsible, Decentralized Intelligence (RDI).
\end{ack}

\bibliographystyle{plainurl} 
\bibliography{refs} 

\end{document}